\documentclass[letterpaper]{article}

\usepackage[square, comma, sort&compress, sectionbib, numbers]{natbib}

\usepackage[nonatbib,final]{nips_2018}

\usepackage[utf8]{inputenc} 
\usepackage[T1]{fontenc}    
\usepackage{hyperref}       
\usepackage{url}            
\usepackage{booktabs}       
\usepackage{amsfonts}       
\usepackage{nicefrac}       
\usepackage{microtype}      
\usepackage{color,soul}
\usepackage{graphicx,graphics,placeins,subcaption,amsmath}
\usepackage{changes}
\usepackage{hyperref}
\usepackage{siunitx}

\title{Approximating the solution to wave propagation using deep neural networks}

\author{
  Wilhelm~E.~Sorteberg
\\
  Dept. of Bioengineering\\
  Imperial College London\\
  SW7 2AZ, UK \\
  \texttt{wilhelm@sorteberg.eu} \\
  \And
  Stef~Garasto \\
  Dept. of Bioengineering\\
  Imperial College London\\
  SW7 2AZ, UK \\
  \texttt{stef.grs@gmail.com} \\
  \And
  Alison~S.~Pouplin \\
  Dept. of Bioengineering\\
  Imperial College London\\
  SW7 2AZ, UK \\
  \And
  Chris~D.~Cantwell \\
  Dept. of Aeronautics\\
  Imperial College London\\
  SW7 2AZ, UK \\
  \And
  Anil~A.~Bharath \\
  Dept. of Bioengineering\\
  Imperial College London\\
  SW7 2AZ, UK \\
}

\begin{document}

\maketitle

\begin{abstract}
Humans gain an implicit understanding of physical laws through observing and interacting with the world. Endowing an autonomous agent with an understanding of physical laws through experience and observation is seldom practical: we should seek alternatives. Fortunately, many of the laws of behaviour of the physical world can be derived from prior knowledge of dynamical systems, expressed through the use of partial differential equations. In this work, we suggest a neural network capable of understanding a specific physical phenomenon: wave propagation in a two-dimensional medium. We define ``understanding'' in this context as the ability to predict the future evolution of the spatial patterns of rendered wave amplitude from a relatively small set of initial observations. The inherent complexity of the wave equations -- together with the existence of reflections and interference -- makes the prediction problem non-trivial. A network capable of making approximate predictions also unlocks the opportunity to speed-up numerical simulations for wave propagation. To this aim, we created a novel dataset of simulated wave motion and built a predictive deep neural network comprising of three main blocks: an encoder, a propagator made by 3 LSTMs, and a decoder. Results show reasonable predictions for as long as 80 time steps into the future on a dataset not seen during training. Furthermore, the network is able to generalize to an initial condition that is qualitatively different from those seen during training\footnote{Accepted to the NeurIPS 2018 Workshop "Modeling the Physical World: Perception, Learning, and Control".}.
\end{abstract}

\section{Introduction}
Predicting the evolution of states in physical systems is a key task of higher intelligence. In engineering, algorithms that implement numerical solvers for modelling and simulation perform the same practical function and are used in many different fields, including aeronautics, mechanical, fluids and electromagnetic systems design. These techniques achieve high accuracy, but can be computationally very expensive; solvers may also require extensive parameter tuning, and the dynamics are fixed rather than learned from data. There has been recent interest in using deep networks and machine learning to replace the role of traditional solvers~\cite{finn2016nips,ehrhardt2017a,sanchez2018,pathak2017,schenck2018spnets}. When used as part of an iterated design workflow, approximate solutions -- generated with lower computational cost than traditional solvers -- could reduce the time needed for design iteration.

Previous uses of deep neural networks to predict the evolution of states of a system can be found within a variety of fields, such as predictions of a robot end-effector's interaction with items in its surrounding \cite{finn2016nips}, trajectories~\cite{ehrhardt2017a}, multi-body dynamics~\cite{sanchez2018}, fluid dynamics~\cite{schenck2018spnets}, and the evolution of chaotic systems~\cite{pathak2017}, among others. A commonly used strategy consists in moving the spatiotemporal data to a latent space, where propagation in time occurs \cite{ehrhardt2017a,finn2016nips,pathak2017}. However, the way the system of interest is given as input to the network can vary. For example, Finn et al. use a guided procedure where the network is presented with sensory data on actuators' live position and forces~\cite{finn2016nips}, Ehrhardt et al. use a neural network which solely relies on visual input~\cite{ehrhardt2017a}, and Sanchez-Gonzales et al. use multiple graph networks containing information about static or dynamic properties of the system~\cite{sanchez2018}.

The main contribution of the present study is the investigation of the use of deep learning to perform state-evolution prediction on the rendered amplitudes of waves propagating through a two-dimensional medium with solid wall boundary conditions. Given the complexity of the equations involved, such a system is non-trivial to solve. Furthermore, the network acts solely on visual input, with no other knowledge of the operating environment. From maps of rendered amplitude observed at 5 instants in time (at the arbitrary but constant sampling rate of 100 Hz), we predict the rendered amplitude patterns over the next 10 time steps (i.e. 100 ms) using an encoder-propagator-decoder structure~\cite{ehrhardt2017a}.  We make the test of prediction quality harsh: the last 5 predictions are repeatedly used as new inputs to enable long term predictions (up to 80 time steps, i.e. 800 ms, into the future). Finally, we test the generalization ability of the network using a qualitatively new initial stimulus pattern.

\section{The Saint-Venant equations}

\begin{figure}[htbp]
\centering
\begin{subfigure}[b]{0.19\linewidth}
\includegraphics[width=0.9\linewidth]{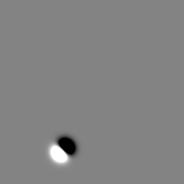}
\caption{$T$ = 0.}
\end{subfigure}
\begin{subfigure}[b]{0.19\linewidth}
\includegraphics[width=0.9\linewidth]{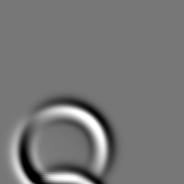}
\caption{$T$ = 25.}
\end{subfigure}
\begin{subfigure}[b]{0.19\linewidth}
\includegraphics[width=0.9\linewidth]{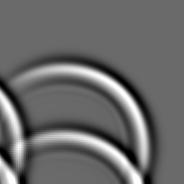}
\caption{$T$ = 50.}
\end{subfigure}
\begin{subfigure}[b]{0.19\linewidth}
\includegraphics[width=0.9\linewidth]{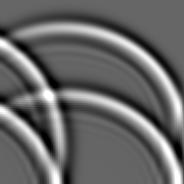}
\caption{$T$ = 75.}
\end{subfigure}
\begin{subfigure}[b]{0.19\linewidth}
\includegraphics[width=0.9\linewidth]{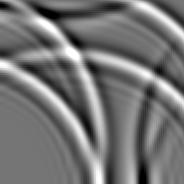}
\caption{$T$ = 100.}
\end{subfigure}

\caption{Illustration of the propagation of one exemplary data point, generated using TriFlow \cite{nicolas_cellier_2018_1239703}. $T$ represents the number of time steps since the initial condition.}
\label{tank_size}
\end{figure}

We aim to solve the 2-dimensional Saint-Venant equations in non-conservative form with no coriolis or viscous forces (a non-linear coupled system), given by:

\begin{eqnarray}
    h_t + ((H+h)u)_x + ((H+h)v)_y & = & 0 \\
    u_t + uu_x + vu_y +gh_x - \nu (u_{xx} + u_{yy}) & = & 0\\
    v_t + uv_x + vv_y +gh_y - \nu (v_{xx} + v_{yy}) & = & 0
\end{eqnarray}
Here, \(u\) and \(v\) are the velocities in the \(x\) and \(y\) direction, respectively, \(H\) is the reference water height and \(h\) is the deviation from this reference, \(g\) is the acceleration due to gravity, \(\nu\) is the kinematic viscosity (set to \(10^{-6}\) \si{m^{2}s^{-1}} ), and we note \(u_x\) the partial derivative of \(u\) with respect to \(x\). Saint-Venant equations are well suited for wave propagation phenomena and are utilized in a variety of fields, including urban flood modelling~\cite{oezgen2016}. The environment was restricted with solid wall boundary conditions along its perimeter. We used a droplet (a localized two-dimensional Gaussian) as the initial condition. The location of the droplet and the propagation speed were allowed to vary within the dataset. The simulated data were produced using TriFlow, a publicly available Python package~\cite{nicolas_cellier_2018_1239703}. The final images are given by a rendering model with azimuth of 45 degrees and altitude of 20 degrees. The generation of each simulated frame took, on average, 0.44 seconds. An example of wave pattern evolution generated by the numerical solver is illustrated in Figure~\ref{tank_size}. The dataset comprised 3,000 simulations, each containing 100 images with resolution \(128 \times 128\) and sampling frequency of 100 Hz (totalling 300,000 images). Random data augmentation was implemented, as well. 

\section{Long term predictor network architecture}
\begin{figure}[htbp]
\centering
\includegraphics[width=.9\linewidth]{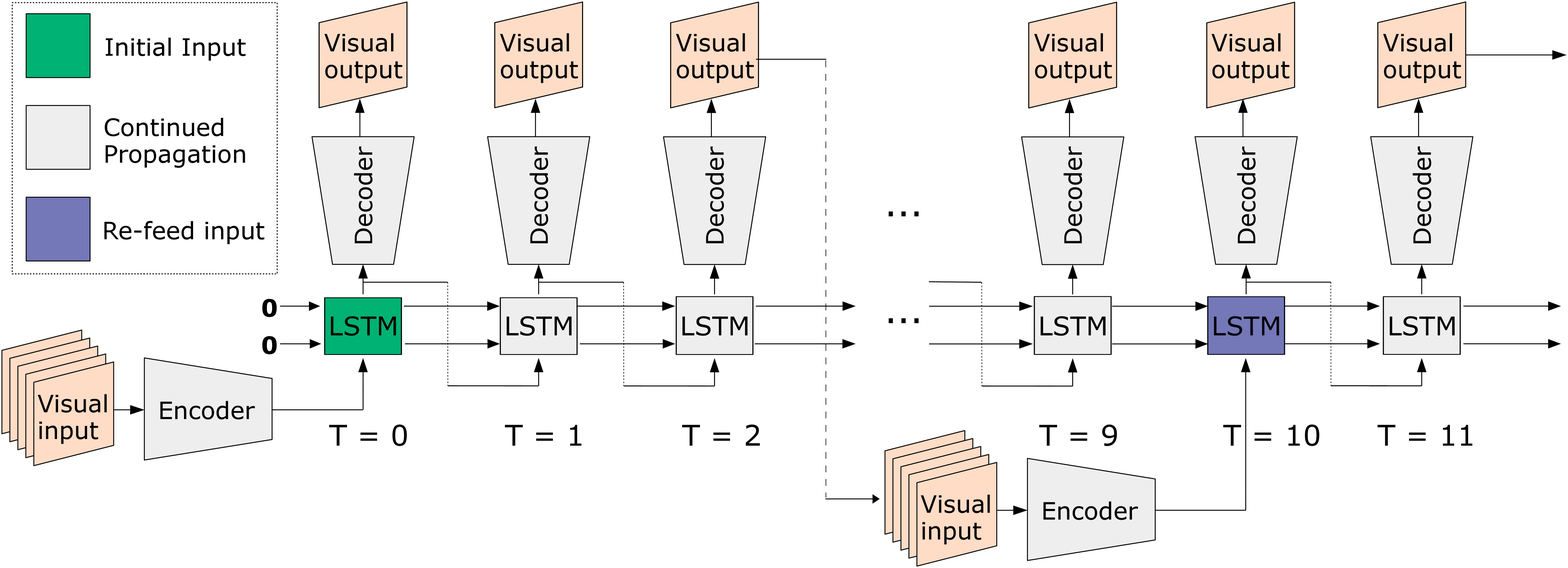}

\caption{Depiction of the internal structure and data flow of the prediction network with 3 LSTMs. Only one of the three LSTMs is used at any time.}
\label{struc2}
\end{figure}

Our neural network consists of an autoencoder structure with a propagator in its centre~\cite{ehrhardt2017a}. The encoder has 4 convolutional layers (60, 120, 240 and 480 units, using a stride of 2 and \(3\times 3\) kernels), and a fully connected layer whose output was a latent space of 1000 nodes. All used the \textit{tanh} non-linearity. Dropout batch normalization layers were used to regularize. After passing the 5 input states (rendered wave patterns) through the encoder, the network propagated the latent space in time repeatedly to generate predictions of 10 future states. From these, the last 5 states were reinserted again as input to propagate further. Propagation in time is made by an elect from 3 fully connected LSTMs, instead of just one like in~\cite{ehrhardt2017a}. The architecture of the decoder mirrors that of the encoder, but with deconvolutional layers. Of the three LSTMs in the propagator only one is active at a time, according to the type of prediction. One LSTM dealt with the initialization of the network, one performed the self-propagation, and the last one was active when predictions were reinserted as a new input. All three LSTMs shared the same hidden and cell states, ensuring complete information preservation (visualized in Figure \ref{struc2}).

We used the Mean Square Error (MSE) between targets and predictions as the loss function, and ADAM as the optimization scheme. At training and test time, 20 and 80 frames were predicted, respectively, thus testing the network on its ability to extrapolate for longer time periods than experienced during training. Using Pytorch and a NVIDIA GTX 1070, convergence was achieved in approximately 7 hours. We evaluated the image quality of the predictions using the Structural Similarity (SSIM) index.

\section{Results and discussion}
\begin{figure}[ht]
\centering
\includegraphics[width=1\linewidth]{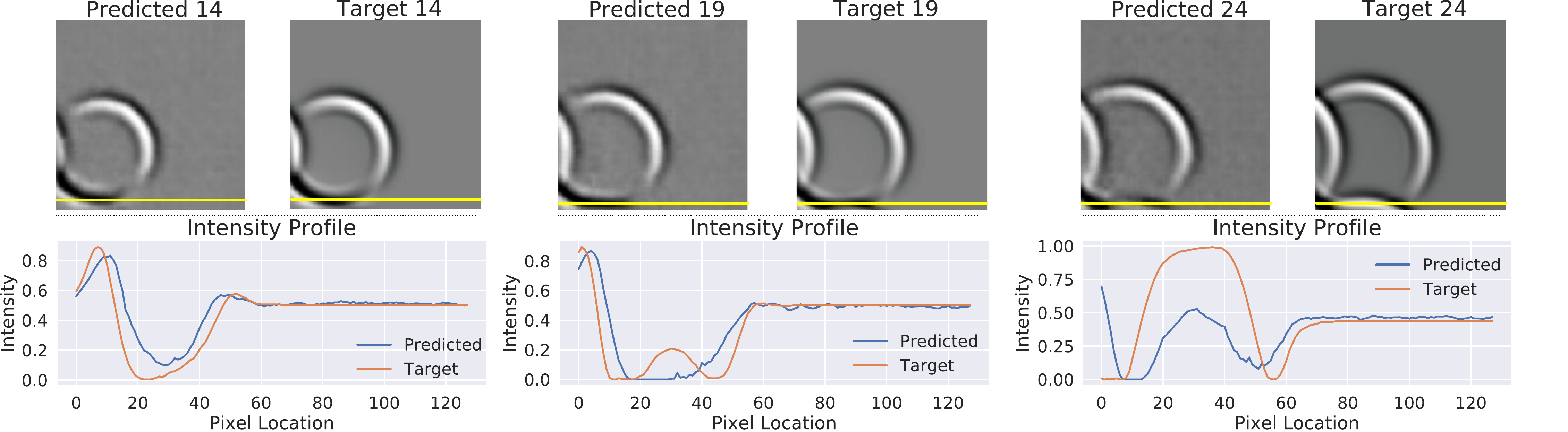}

\caption{Top: predicted two-dimensional spatial patterns vs targets from numerical solver. Bottom: spatial intensity profiles for predictions and targets along the horizontal line in the figures above.}
\label{cutthrough}
\end{figure}

\begin{figure}[htbp]
\begin{subfigure}[b]{.38\textwidth}
\includegraphics[width=.95\textwidth,height=0.52\textwidth]{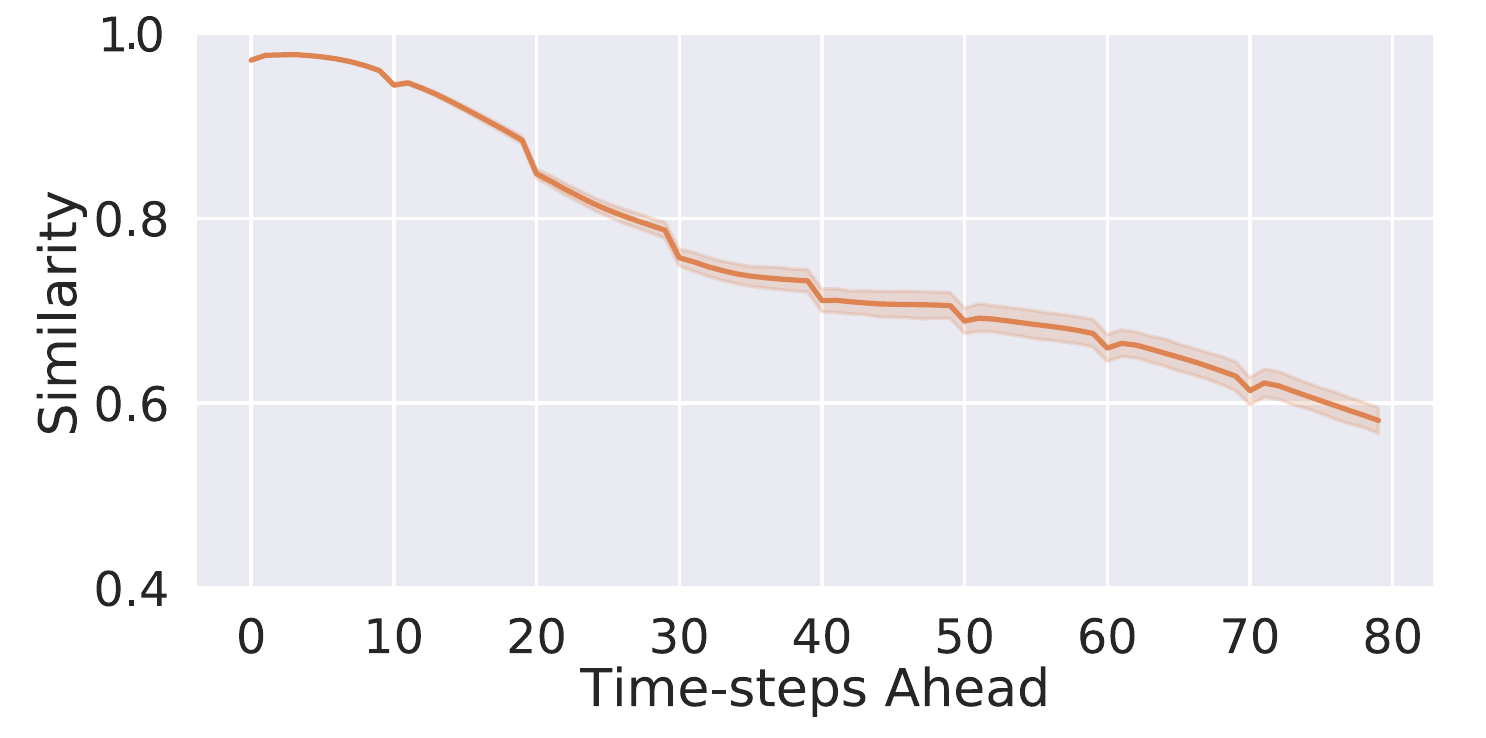}
\caption{}
\end{subfigure}
\begin{subfigure}[b]{.58\textwidth}
\centering
\includegraphics[width=.95\textwidth]{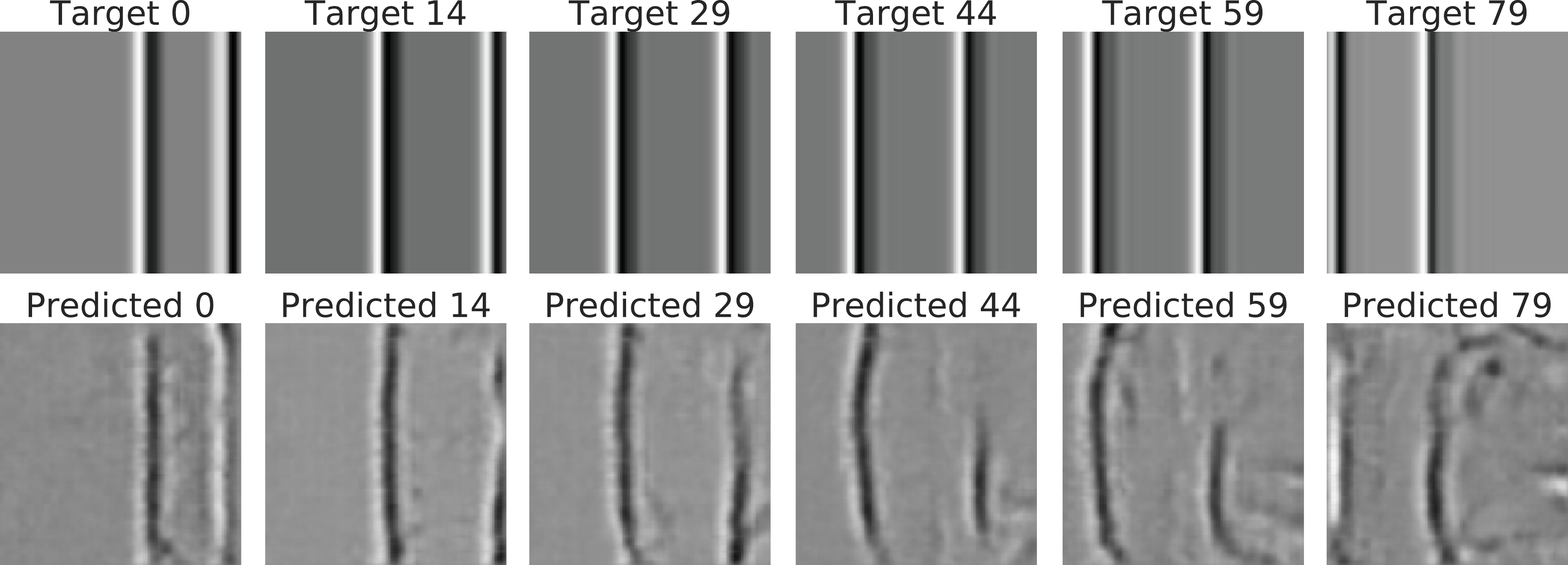}
\caption{}
\end{subfigure}

\caption{(a) SSIM between targets and predictions vs.\ time-steps into the future of the prediction. (b) A comparison of the predicted vs.\ target frames at six selected time-steps using a linear wave front as initial excitation. Though coherence begins to deteriorate, wave-speed is largely preserved.}
\label{prediction_Line}
\end{figure}

Some examples of target and predicted rendered wave patterns can be seen in Figure \ref{cutthrough}, together with a profile along the row of maximum variance of the ground truth image. It can be seen that the network is able to replicate the spatial pattern of the rendered wave amplitude (top row), albeit with lower accuracy around reflections at the border (bottom row). Figure \ref{prediction_Line}a, instead, shows the SSIM between targets and predictions versus the time at which predictions were made. The solid line and the shaded area are, respectively, the mean and the $95\%$ confidence interval (CI) across the test dataset. As expected, the SSIM decreases for longer term predictions, but the contained increase of the CI is promising.

On top of assessing the network on left-out test data, we also tested it on a qualitatively different type of initial condition, specifically a propagating linear wave front. The results are shown in Figure \ref{prediction_Line}b. Although these predictions are not as accurate as those of Figure~\ref{cutthrough}, they suggest that the network is able to generalize on completely unseen data which varies significantly in spatial distribution from the training data. However, the influence of the circular wave front -- as a prior in the training data -- is also clearly visible, especially for predictions further ahead in time. More example results can be found \href{http://bit.ly/2PBvjms}{here}.\footnote{http://bit.ly/2PBvjms}

The results suggest that our trained network can perform implicit understanding and state prediction in environments governed by complex physical laws. From a five-frames visual input, the network can predict up to 80 frames with good accuracy, albeit with  lower quality than original numerical solver. However, the network predictions take less than half a second (for 80 frames) -- as opposed to the 35 seconds (on average) employed by the numerical algorithm. With the proviso of sufficiently general training, similar networks could be used as an approximate simulation tool, to quickly iterate over designs, or perform initial searches through parameter space. Results on the generalization to unseen initial conditions are also promising. For future work, we plan to increase prediction accuracy, to test the neural network on new generalization scenarios, to analyze the representation it learns in latent space, and to extend this study to data from real-world wave propagation.

\medskip

\small


\end{document}